\documentclass[onecolumn,floats,superscriptaddress,showpacs]{revtex4}
\usepackage{amssymb}

\usepackage{graphicx}

\begin{document}

\title{Inhomogeneous states in non-equilibrium superconductor/normal metal
tunnel structures: a LOFF-like phase for non-magnetic systems.}
\author{A. Moor$^{1}$, A.F. Volkov$^{1,2}$, and K. B. Efetov.}

\address{Theoretische Physik III,\\
Ruhr-Universit\"{a}t Bochum, D-44780 Bochum, Germany\\
$^{2}$Institute for Radioengineering and Electronics of Russian Academy of\\
Sciences,11-7 Mokhovaya str., Moscow 125009, Russia\\}

\begin{abstract}
We analyze non-equilibrium states in a tunnel superconductor-normal metal
(N/S/N) structure in the presence of a tunnel current $I$. We use an
approximation of an effective temperature $T$ and calculate the
current-voltage I-V characteristics. It is shown that the I-V dependence may
have an S-shaped form. We determine nonuniform current $I(x)$ and
temperature $T(x)$ distributions that arise as a result of instability of
the uniform state with negative differential conductance ($dI/dV<0$). We
discuss an analogy with equilibrium superconductors with an exchange field
in which nonuniform states predicted by Larkin/Ovchinnikov and Fulde/Ferrell
are possible.
\end{abstract}

\maketitle

\bigskip

\section{Introduction}

Study of superconductor/normal metal (S/N) or superconductor/superconductor
hetero-structures has been a popular topic during last decades \cite%
{LambRaim,Been,NazRev,ZaikinRev}. Both the equilibrium and non-equilibrium
properties of such systems were under intensive investigation.
Non-equilibrium phenomena are of special interest because they are
essentially different from the equilibrium ones but several interesting
effects have been observed even in the cases when a deviation from the
equilibrium is small. As an example, one can mention a non-monotonic
behavior of the resistance of S/N structures as a function of temperature $T$
or applied voltage $V$ \cite{AVZ,Marg,NazarovPRL,LambVolkov,Pannetier}.

If the applied voltage is not too small, the superconductor S (for example,
in a N/S/N system) goes out of the equilibrium and new phenomena come into
play. Study of non-equilibrium effects in superconducting hetero-structures
as well as in homogeneous superconductors in the presence of optical or
microwaves radiations has been carried out during a long time (see reviews %
\cite{Elesin,SchonRev,Mints} and references therein). For example,
Eliashberg suggested to stimulate superconductivity by a microwave radiation
that leads to an essential deviation of the distribution function $f$ from
the equilibrium one \cite{Eliashberg}. As the energy gap $\Delta $ in S is
related to the function $f$ via the self-consistancy equation, the energy
gap and the critical temperature may increase in the presence of \ a
microwave radiation, which results in an increase of the critical current %
\cite{Ivlev}. Another interesting effect caused by a non-equilibrium
distribution function is the sign reversal of the Josephson critical current
$j_{c}$ in a multi-terminal SNS junction \cite{KlapWees}. It turns out that,
provided the distribution function in the normal metal N controlled by
additional N' electrodes differs significantly from the equilibrium one, the
critical current $j_{c}$ becomes negative ($\pi $-state) \cite%
{VolkovPRL,SchonPRL,Yip}.

In recent years, the interest in studies of non-equilibrium effects in S/N
or S/S' hetero-structures has revived. This is due to the progress in
nanotechnology, the possible applications of such structures in
low-temperature devices \cite{Pekola} and to the progress in theoretical
research \cite{LambRaim,Been,ZaikinRev,NazRev,VZK}.

Perhaps, the simplest system in which one can study non-equilibrium effects
is a N/S/N structure with a bias current $I_{b}$. Such a structure was
studied theoretically in recent papers \cite{Keizer,Nazarov}. In Ref.\cite%
{Keizer} it was assumed that the resistance of the S/N interfaces is
negligible. The calculated I-V characteristics (CVC) was shown to be of the
so-called N-shape type, that is, three values of voltages correspond to one
value of the bias current $I_{b}$. Earlier this type of the I-V
characteristics was studied in other superconducting systems \cite%
{Tinkham,VKJETP,Mints,Vodolazov}.

The opposite case of a large S/N interface resistance was considered in Ref.%
\cite{Nazarov}. In this case the I-V curve is of the S-type, i.e.
three values of the current correspond to one value of the voltage $V$
between the N leads. Such kind of the CVC was known from previous studies of
the superconducting structures. For example, it can be realized in S/I/S
junctions \cite{Dynes,Gray,Elesin,SchonRev} and in granular superconductors %
\cite{India,Al'tshuler}. The authors of Ref.\cite{Nazarov} investigated the
stability of this system and came to the conclusion that the system is
stable. However, this conclusion is valid only in the case of small lateral
dimensions of the structure. From a general theory of the system with
negative differential conductance ($G_{d}=dI/dV<0$), it is known that the
states with negative $G_{d}$ are unstable and in case of the S-shape I-V
curve a stratification of the current density occurs as a result of the
instability \cite{Readley,VK}. Inhomogeneous states in different
superconducting systems were studied in approximate models with account for
electron-electron or electron-phonon inelastic scattering \cite%
{Elesin,SchonRev}.

In the present paper, we analyze the N/S/N structure of the type considered
in Ref.\cite{Nazarov} assuming however that lateral dimensions are large.
The latter assumption turns out to be very important. We find in this limit
that a new inhomogeneous state is possible, which contrasts the situation in
small sized systems. Physics of this inhomogeneous state is very similar to
that of the inhomogeneous state in equilibrium superconductors with an
exchange field (the Larkin-Ovchinnikov-Fulde-Ferell states \cite{LO,FF}). It is worth emphasizing
though that, in the system considered here, there are no magnetic
interactions and their role is played by a finite voltage $V$ (see below).\
Under certain assumptions we calculate the I-V curve and study possible
inhomogeneous states. Most of the results are obtained in an analytical
form. In particular, we find the spatial distribution of the current density
$j(x)$ and energy gap $\Delta (x)$ and draw an analogy with those in the
LOFF states.

\bigskip

\section{Model. Basic Equations}

In order to calculate physical quantities we use a kinetic equation for the
non-equilibrium distribution functions $f_{\pm }$. The procedure of the
derivation is quite standard (see, e.g. \cite{Kopnin}) and is based on using
quasi-classical Keldysh matrix Green functions $\hat{g}$. They are expressed
in terms of the two distribution functions $f_{\pm }$ as $\hat{g}=\hat{g}%
^{R}(\hat{\tau}_{3}f_{-}+\hat{\tau}_{0}f_{+})-(\hat{\tau}_{3}f_{-}+\hat{\tau}%
_{0}f_{+})\hat{g}^{A}$, where $\hat{\tau}_{3}$ is the $\hat{\tau}_{z}$ Pauli
matrix and $\hat{\tau}_{0}$ is the unit matrix. The functions $f_{+}$ and $%
f_{-}$ correspond to the functions $f_{L}$ and $f_{T}$ introduced by Schmid
and Sch\"{o}n \cite{Schmid}.

The considered N/S/N system is shown in Fig.1a. We assume that the
tunnelling probability through the interface is small and that the retarded
(advanced) Green's functions $\hat{g}^{R(A)}$ have the same form as in a
bulk superconductor: $\hat{g}^{R(A)}=g^{R(A)}\hat{\tau}_{3}+f^{R(A)}i\hat{%
\tau}_{2}$ with $f^{R(A)}=\Delta /\sqrt{(\epsilon \pm i\gamma )^{2}-\Delta
^{2}}$ and $g^{R(A)}=\sqrt{1-(f^{R(A)})^{2}}$. The only difference is that,
in contrast to the bulk case, the damping $\gamma $\, caused by a weak
proximity effect, enters the formula for the function $f^{R(A)}$. In this
limit the damping $\gamma $ is equal to $\gamma \approx \epsilon _{0},$
where $\epsilon _{0}=D/(R_{SN}\sigma d)$, $D=v^{2}\tau /3$ is the diffusion
coefficient, $R_{SN}$ is the SN interface resistance per unit area, $\sigma $
and $d$ are the conductivity (in the normal state) and the thickness of the
superconductor. The thickness of the S layer is assumed to be small ($d<<\xi
_{S}\approx \sqrt{D/\Delta }$), so that all quantities do not depend on the $%
z-$coordinate. The opposite case was considered in Ref. \cite{Keizer,Hekking}%
. The thickness of the N leads is supposed to be large ($d_{N}>>\xi _{S}$)
so that the N metals are in the equilibrium state.

\begin{figure}[tbp]
\begin{center}
\includegraphics[width=8cm, height=10cm]{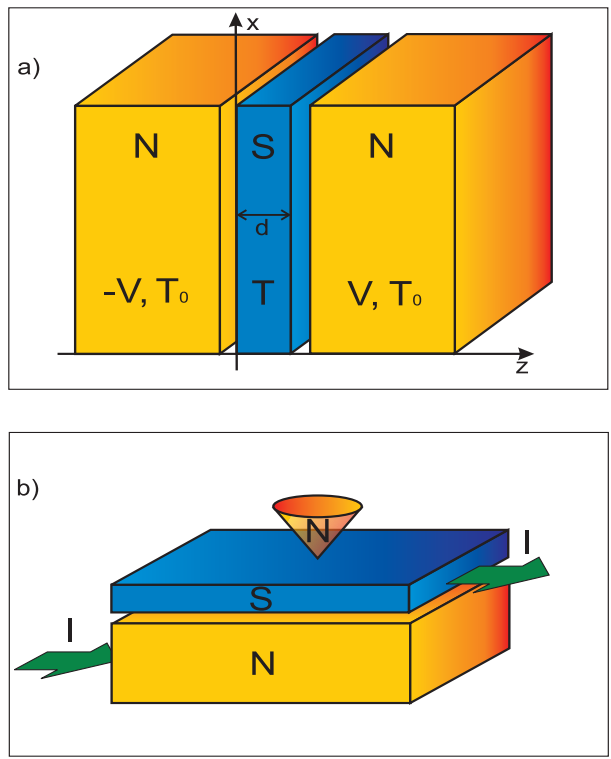}
\end{center}
\caption{a) The system under consideration. b) The proposed system in which
an inhomogeneous distribution of the energy gap can be probed with a point
contact spectroscopy.}
\end{figure}

The function $f_{-}\left( \epsilon ,V\right) $ appears due to a branch
imbalance and, being dependent on the electric potential $V$, determines the
current $I$. In case of a fully symmetric N/S/N system $f_{-}=0$. The
function $f_{+}\left( \epsilon ,V\right) $ determines the energy gap $\Delta
$. Since the function $f_{+}\left( \epsilon ,V\right) $ depends on the
applied voltage $V,$ the gap $\Delta $ is also a function of $V$. The
dependence $\Delta (V)$ is given by the self-consistency equation

\begin{equation}
1=\lambda Re\int_{0}^{\theta _{D}}d\epsilon \frac{f_{+}(\epsilon ,V)}{\sqrt{%
\tilde{\epsilon}^{2}-\Delta ^{2}}}  \label{SCeq}
\end{equation}%
where $\tilde{\epsilon}=(\epsilon +i\gamma ),$ $\lambda $ and $\theta _{D}$
are the coupling constant and Debye energy, respectively. Eq. (\ref{SCeq})
is a generalization of the conventional BCS equation to the non-equilibrium
case.

The functions $f_{\pm }$ can be found from the kinetic equation that can be
written in S in the form (see for example \cite{SV})

\begin{equation}
\frac{\partial (\mathcal{N}_{S}(\epsilon )f_{\pm })}{\partial t}-\frac{%
\partial (\mathcal{D}_{\pm }(\epsilon )\partial f_{\pm })}{\partial x^{2}}%
=\sum_{\alpha =l,r}\epsilon _{\alpha }\mathcal{A}_{\alpha \pm }(\epsilon
)+S_{in}\{f_{\pm }\}  \label{KineticEq}
\end{equation}%
where
\begin{equation}
\mathcal{N}_{S}(\epsilon )=Re\{g^{R}(\epsilon )\}=Re\frac{\tilde{\epsilon}}{%
\sqrt{\tilde{\epsilon}^{2}-\Delta ^{2}}}  \label{a0}
\end{equation}%
is the normalized density-of-states (DOS), $\epsilon _{l,r}=(D/R_{l,r}\sigma
d)$ with $R_{l,r}$ being the resistance of the left (right) S/N interfaces,
and $\mathcal{D}_{\pm }(\epsilon )$ is an energy-dependent diffusion coefficient,
\begin{equation}
\mathcal{D}_{\pm }(\epsilon )=\frac{D}{2}\left[ 1+\frac{\tilde{\epsilon}%
\tilde{\epsilon}^{\ast }\mp \Delta ^{2}}{\sqrt{\tilde{\epsilon}^{2}-\Delta
^{2}}\sqrt{\tilde{\epsilon}^{\ast 2}-\Delta ^{2}}}\right]  \label{a1}
\end{equation}%
with $\tilde{\epsilon}^{\ast }=\epsilon -i\gamma $.

The functions $\mathcal{A}_{l,r\pm }(\epsilon )$ in Eq. (\ref{KineticEq})
are defined as:
\begin{equation}
\mathcal{A}_{l,r\pm }(\epsilon )=\mathcal{N}_{S}(\epsilon )[F_{\pm
}(\epsilon ,V_{l,r})-f_{\pm }(\epsilon )]  \label{a2}
\end{equation}

The distribution functions in the normal metals are assumed to have the
equilibrium forms shifted by the applied voltages $V$. In a fully symmetric
system, the voltages in the right ($V_{r}$) and left ($V_{l}$) N metals are
equal to $V_{r}=-V_{l}\equiv V$. In this situation, the functions $F_{\pm } $
entering Eq. (\ref{a2}) take the simple form
\begin{equation}
F_{\pm }=(\tanh (\epsilon +eV)\beta _{0}\pm \mathbf{\ }\tanh (\epsilon
-eV)\beta _{0})/2  \label{a3}
\end{equation}%
where $1/2\beta _{0}=T_{0}$\ is the reservoir temperature. In the limit of
the weak coupling between N and S layers the density-of-states in the N
metals is supposed to be unperturbed by the proximity effect: $\mathcal{N}%
_{N}(\epsilon )=1.$

The first terms in the R.H.S. of the kinetic equation (\ref{KineticEq})
describe the tunneling of quasiparticles from (to) the normal electrodes,
whereas the last one is an inelastic collision term. It consists of the
electron-electron and electron-phonon collision terms ($%
S_{in}=S_{e-e}+S_{e-ph}$) and is of the order of $\ f_{\pm }(\epsilon )/\tau
_{e-e},f_{\pm }(\epsilon )/\tau _{e-ph},$ where $\tau _{e-e}$ and $\tau
_{e-ph}$ are the electron-electron and electron-phonon inelastic scattering
times. Eq.(\ref{KineticEq}) contains the DOS, $N_{S}(\epsilon ,\Delta )$,
depending on the energy gap $\Delta (V).$ Therefore Eqs.(\ref{SCeq}-\ref{a3}%
) are coupled, and the problem of finding the distribution function $f_{+}$
is not easy.

\bigskip

\section{Moderate Interface Resistance. Homogeneous Case}

We consider first a homogeneous stationary state assuming that all the
quantities do not depend on the coordinate $x$. However, the problem of
solving Eqs.(\ref{SCeq}-\ref{a3}) is too complicated even under this
assumption and therefore we consider only two limiting cases.

First, we analyse the limit of a very weak electron-electron and
electron-phonon interactions:

$\Delta >>\epsilon _{l,r}>>\tau _{e-e}^{-1},\tau _{e-ph}^{-1}$, when the
collision terms in Eq. (\ref{KineticEq}) can be neglected. In other words,
the S/N interface resistance $R_{SN}$ is not too high: $R_{SN}\sigma
d<<\{D\tau _{e-e},D\tau _{e-ph}\}$. This approximation has been used in many
papers (see, for instance, \cite{VZK,Nazarov}). For a symmetric N/S/N system
we use Eq. (\ref{a3}) and obtain from Eq.(\ref{KineticEq}) the following
solutions
\begin{equation}
f_{+}=F_{+},\quad f_{-}=0  \label{a4}
\end{equation}%
Substituting the function $f_{+}$ into Eq.(\ref{SCeq}) and shifting the
energy in the integral by $eV$ in the first term and by $-eV$ in the second
one, we reduce Eq. (\ref{SCeq}) to the following form

\begin{equation}
1=\frac{\lambda }{2}Re\int_{-\theta _{D}}^{\theta _{D}}\frac{\tanh \left(
\epsilon \beta _{0}\right) }{\sqrt{(\epsilon -eV+i\gamma )^{2}-\Delta ^{2}}}%
d\epsilon  \label{a5}
\end{equation}%
Remarkably, the form of Eq. (\ref{a5}) is the same as the one for a
superconductor in the presence of an exchange field. In other words, the
problem of the non-equilibrium superconductor in the N/S/N system is to a
great extent equivalent to the problem of the equilibrium superconductor
with the distribution function $f_{+}=\tanh (\epsilon \beta _{0})$ in the
presence of an ``exchange''\ field $eV$.

The latter problem has been attracting a lot of attention since the
pioneering works by Larkin and Ovchinnikov \cite{LO} and by Fulde and Ferrell %
\cite{FF}. For the model with the exchange field, these authors have
predicted a new state called now LOFF state. They considered a clean
superconductor with the exchange field $h$ (or a strong magnetic field)
acting on spins of electrons. They demonstrated that at zero temperature the
energy gap $\Delta $ remained unchanged unless the exchange energy $h$
exceeded the value $\Delta _{0}$, where $\Delta _{0}$ is the energy gap at
zero temperature in the absence of $h$.

However, in addition to this solution of the self-consistency equation,
there is another, unstable, solution for $\Delta :\Delta (h)=\Delta _{0}\sqrt{h^{2}/h_{c}^{2}-1}$
with $h_{c}^{2}=\Delta _{0}^{2}/2$ and $h_{c}\leq h\leq \Delta _{0}.$ Therefore, in
the interval $h_{c}\leq h\leq \Delta _{0}$ there are three possible
solutions for $\Delta :0$, $\Delta (h)$ and $\Delta _{0}$ (the trivial
solution $\Delta =0$ always exists). The situation resembles the behavior of
a non-ideal gas described by the van der Waals equation of state, and thus,
one can expect a stratification of the electron system. Indeed, as it was
shown in Refs. \cite{FF,LO}, an inhomogeneous state with the energy gap $%
\Delta (r)$ varying in space turns out to be more favorable than the
homogeneous one.

Using the equivalence between the N/S/N system at a finite voltage $V$ and
the equilibrium superconductor in the presence of the exchange field one may
expect an inhomogeneous LOFF-like state in the system considered here. A
multi-valued (in a certain region of parameters) dependence of the energy
gap $\Delta $ on the applied voltage $V$ and damping $\gamma $ has recently
been found for an N/S/N system in Ref. \cite{Nazarov}. The function $\Delta
(V)$ had a form similar to that determined by Eq. (\ref{a5}). It was
established that at some values of the parameters the CVC had the so-called
S-shaped form. This type of the CVC is well known in bulk superconductors
and has already been obtained both theoretically (see the review \cite%
{Elesin,SchonRev}) and experimentally \cite{Dynes,Gray} several decades ago.

Although, one could expect an inhomogeneous state in the N/S/N systems, the
authors of Ref. \cite{Nazarov} came to the conclusion that the homogeneous
state had to remain stable. This statement is correct for the superconductor
island of a small size considered in Ref. \cite{Nazarov} but cannot remain
valid for large superconducting films sandwiched between normal metals. Note
also that the authors of Ref.\cite{Nazarov} studied the stability of the
system in a short time interval: $t\lesssim \hbar /\Delta $ (a similar
dynamic behavior of the order parameter in a collisionless superconducting
system was studied earlier in Ref. \cite{VK73}). On the other hand the
instability develops on much longer characteristic times $t\sim \hbar
/\epsilon _{0}.$

In this case, the states corresponding to the part of the CVC with negative
differential resistance are unstable and the stratification of the current
density occurs in the system (see the review \cite{VK}). This structure
resembles the LOFF coordinate dependence of the order parameter and of other
physical quantities. However, in contrast to the LOFF state in the
equilibrium superconductors with the exchange field, the coordinate
dependence cannot be found minimizing the free energy because we consider a
system out of the equilibrium.

\begin{figure}[tbp]
\begin{center}
\includegraphics[width=6cm, height=9cm]{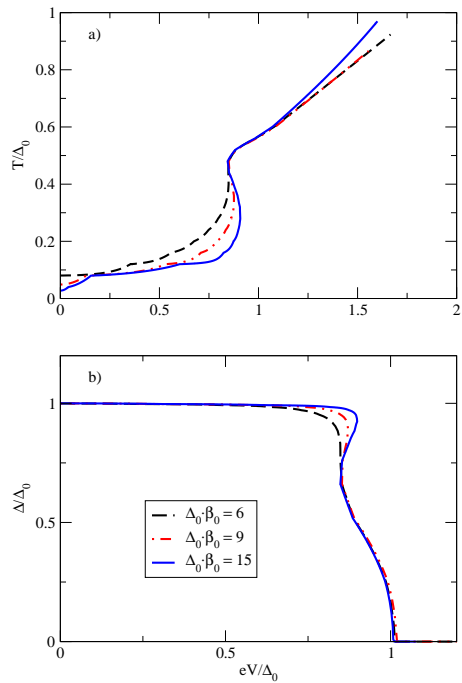}
\end{center}
\caption{Normalized effective temperature (a) and energy gap (b) as
functions of the applied voltage $V$ for $\protect\gamma /\Delta _{0}=0.01$
and different temperatures $T_{0} = 1/(2\protect\beta_{0})$ (shown in the
inset)}
\end{figure}

\bigskip

\section{High Interface Resistance. Non-homogeneous Case}

Although the similarity between the problem involved and the problem of LOFF
is most clearly seen in the limit, $\epsilon _{l,r}>>\tau _{e-e}^{-1},\tau
_{e-ph}^{-1}$, one has to solve in this limit complicated
integro-differential equations (\ref{SCeq},\ref{KineticEq}) and this can be
done only numerically. The problem can be solved analytically in another
limit:

\begin{equation}
\tau _{e-ph}^{-1}<<\epsilon _{l,r}<<\tau _{e-e}^{-1}  \label{EffTcond}
\end{equation}%
In this limit the electron-phonon interaction is very weak and the
electron-phonon collision term in Eq. (\ref{KineticEq}) can be neglected.
The largest term in Eq. (\ref{KineticEq}) is the electron-electron collision
term $S_{e-e}\{f_{\pm }\}.$

Note that the electron-electron $\tau _{e-e}^{-1}$\ and electron-phonon $%
\tau _{e-ph}^{-1}$\ scattering rates depend on energy $\epsilon $ \cite%
{Langenberg} so that Eq.(\ref{EffTcond})\ \ can be satisfied not at all
energies. If the mean free path is not too short ($s/T<l,$ but $l<v/3\Delta $
because we consider the dirty limit), these scattering rates are: $\tau
_{e-e}^{-1}\sim \gamma _{e-e}(T_{e})\epsilon ^{2}/T_{e}^{2}$\ and $\tau
_{e-ph}^{-1}\sim \gamma _{e-ph}(T_{e})\epsilon ^{3}/T_{e}^{3}$\ with $\gamma
_{e-e}=\pi \lambda _{e-e}T_{e}^{2}/(8\hbar E_{F})$\ and $\gamma _{e-ph}=\pi
\lambda _{e-ph}T_{e}^{3}/(2\hbar s^{2}p_{F}^{2}),$\ where $\lambda
_{e-e},\lambda _{e-ph}$\ are electron-electron (electron-phonon) interaction
constants, $E_{F}$\ is the Fermi energy, $s$\ is the sound velocity (see,
for example, \cite{Kopnin}). The characteristic energy $\epsilon _{ch}$ in
our case is of the order $\epsilon _{ch}\sim T_{e}\sim T_{c};$\ thus, for
the Al S-films the inelastic scattering rates are: $\tau _{e-e}^{-1}\approx
10^{8}s^{-1}$\ and $\tau _{e-ph}^{-1}\approx 10^{6}s^{-1}$ \cite{Kopnin1}.\
At temperatures lower than $T_{c}$,\ the interval between $\tau _{e-e}^{-1}$%
\ and $\tau _{e-ph}^{-1}$\ becomes larger because $\tau _{e-ph}^{-1}$\ has a
stronger dependence on $\epsilon $\ than $\tau _{e-e}^{-1}.$\ However, one
has to take into account the dependence of $\tau _{e-e}^{-1}$ and $\tau
_{e-ph}^{-1}$ on $\Delta (T).$ Therefore, if the frequency $\epsilon
_{0}/\hbar =D/(R_{SN}\sigma d),$ which is determined by the S/N interface
resistance $R_{SN},$\ is chosen in the interval $10^{6}s^{-1}$ $-$\ $%
10^{8}s^{-1}$, the approximation of an effective temperature can be applied
to the case of these films.

It is worth noting that the violation of the condition (\ref{EffTcond})
leads only to breakdown of the effective temperature approximation, but not
to disappearance of the effects under consideration. For these effects it is
important only to have the S-shape CVC, which is realized, for example, also
in the case of the frequency $\epsilon _{0}/\hbar $\ large in comparison
with $\tau _{e-e}^{-1}$\ and $\tau _{e-ph}^{-1}$\ (see the preceding section
and \cite{Nazarov}). However the approximation of the effective temperature
allows one to solve the problem in an analytical form. In the limit of high
impurity concentration ($l<s/T$), the inelastic electron-phonon scattering
rate $\tau _{e-ph}^{-1}$ depends on energy even stronger \cite%
{Al'tshuler,Al'tshuler1,Reizer,Sergeev}, that is, the window between $\tau
_{e-e}^{-1}$ and $\tau _{e-ph}^{-1}$ becomes larger and the situation is
even more favorable for applicability of our approach. If the frequency $%
\epsilon _{0}/\hbar $ is less than $\tau _{e-ph}^{-1}$, then the heat
absorbed by quasiparticles in the superconductor is released mainly to the
lattice, but not to the normal electrodes, and our approach is not
applicable. However, one can show that in this case the CVC of the system
also has the S-shape form and all the effects considered here remain
qualitatively unchanged, although the analytical approach is not possible in
this case.

In the limit determined by Eq. (\ref{EffTcond}) the electron-phonon
interaction is very weak and the electron-phonon collision term in Eq. (\ref%
{KineticEq}) can be neglected. The largest term in Eq. (\ref{KineticEq}) is
the electron-electron collision term $S_{e-e}\{f_{\pm }\}.$ Due to a high
rate of the electron-electron collisions one can assume that the
distribution function $f_{+}$ has an equilibrium form, $f_{+}(\epsilon
)=\tanh (\epsilon \beta )$, with an effective temperature $T\equiv (2\beta
)^{-1}$ depending on the coordinate $x$. Perhaps for the first time, the
approximation of the effective temperature has been introduced in the study
of nonequilibrium superconductors by Parker \cite{Parker} (see also \cite%
{VK,Al'tshuler,Foot1}). In order to find this temperature, we multiply Eq. (%
\ref{KineticEq}) by $\epsilon $, drop the first term in the L.H.S. and
integrate the equation over all energies. This procedure leads to an
equation describing the energy conservation. The contribution of the
collision term $S_{e-e}\{f_{+}\}$ equals zero because this term conserves
the total energy, and we obtain the equation for $T$

\begin{equation}
-\frac{\partial (\mathcal{M}(\tilde{T})\partial \tilde{T}/\partial x)}{%
\partial x}=\sum_{\alpha =l,r}\kappa _{\alpha }^{2}[\mathcal{J}(V_{\alpha },%
\tilde{T})-\mathcal{S}(\tilde{T})]  \label{EqTeff}
\end{equation}%
where $\kappa _{l,r}^{2}=(D/R_{l,r}\sigma d),$ $\tilde{T}\equiv T/T_{0}$ is
the normalized temperature, and
\[
\mathcal{M}(\tilde{T})=(\tilde{T}/D)\int_{0}^{\infty }dyy^{2}\mathcal{D}%
_{+}(y,\Delta \beta )\cosh ^{-2}(y).
\]%
where the function $\mathcal{D}_{+}(y,\Delta \beta )$ is defined in Eq.(\ref%
{a1}).

The functions $\mathcal{J}(V_{\alpha },T)$ and $\mathcal{S}(T)$ are the heat
source\ (Joule heat) and drain due to the tunnelling of electrons into the N
electrodes. They can be written as%
\begin{equation}
\mathcal{J}(V_{a},\tilde{T})=\int_{0}^{\infty }dyy\mathcal{N}_{S}(y)[\tanh
(y)-(\tanh (y_{+})+\tanh (y_{-}))/2],  \label{a7}
\end{equation}%
\begin{equation}
\mathcal{S}(\tilde{T})=\int_{0}^{\infty }dyy\mathcal{N}_{S}(y)[\tanh
(y)-\tanh (y/\tilde{T})],  \label{a8}
\end{equation}%
where $y=\epsilon \beta _{0},$ $y_{\pm }=y\pm eV\beta _{0},$ and $\tilde{T}%
=T/T_{0}.$

In order to solve Eq. (\ref{EqTeff}) we consider for simplicity the
symmetric case when $V_{r}=-V_{l}\equiv V.$ In the homogeneous case the
dependence of the effective temperature on the applied voltage can be found
from the balance equation: $\mathcal{J}(V,\tilde{T})=\mathcal{S}(\tilde{T}).$
Explicit expressions can be derived analytically in the limits of low and
high temperatures.

In the limit of high temperatures, $V\beta _{0}<<1,$ we come to the
following formula
\begin{equation}
\tilde{T}-1\equiv (T-T_{0})/T_{0}\approx (V\beta _{0})^{2}  \label{a9}
\end{equation}%
whereas at low temperatures, $V\beta _{0}>>1,|eV-\Delta |>>T_{0}$\ we obtain
\begin{equation}
\tilde{T}-1\approx eV/(\Delta -eV)  \label{a10}
\end{equation}%
Note that at low temperatures $T_{0}$ both the terms $J$ and $S$ are, as
they should be, exponentially small but the effective temperature $T$ is
not. Making these estimations, we assume that $\gamma \rightarrow 0$. Thus,
it is not obvious that the effective temperature $T$ and energy gap $\Delta $
will be multi-valued functions of $V$ as it takes place in another limit
considered in the previous section. Numerical calculations of the integrals (%
\ref{a7},\ref{a8}) show that the situation qualitatively remains unchanged,
i.e. the quantities $T$ and $\Delta $ have three values in a certain
interval of voltages $V$. The dependencies $T(V)$ and $\Delta (V)$ are shown
in Fig.2a and 2b. One can see that in a narrow interval of the voltage, the
effective temperature can have three values ($T_{1}, T_{2}, T_{3}$), and
therefore the energy gap $\Delta $ is also a multi-valued function of $V$,
which is analogous to the behavior in superconductors with an exchange field
$h$.

The current through the system is given by the formula

\begin{equation}
I=\frac{1}{eR_{b}}\int_{0}^{\infty }d\epsilon \mathcal{N}_{S}(\epsilon
,\Delta (T))F_{-}(\epsilon ,V)  \label{Current}
\end{equation}%
where $F_{-}=(\tanh (\epsilon +eV)\beta _{0}-$ $\tanh (\epsilon -eV)\beta
_{0})/2.$ The CVC obtained from Eq. (\ref{Current}) is shown in Fig.3 for
different temperatures $T_{0}$ (Fig.3a) and damping $\gamma$ (Fig.3b).

\begin{figure}[tbp]
\begin{center}
\includegraphics[width=5cm, height=8cm]{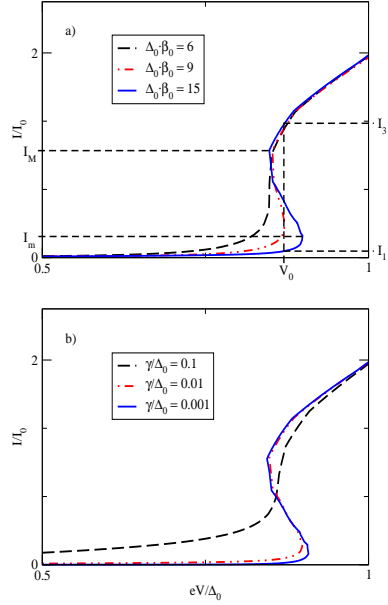}
\end{center}
\caption{ a) The CVC of the system in the homogeneous state for $\protect%
\gamma /\Delta _{0}=0.01$ and different temperatures. The part of the CVC
corresponding to an inhomogeneous state (a broad high temperature domain) is
shown by the dashed vertical line. b) The CVC in the homogeneous state for
different damping $\protect\gamma$. The temperature ${T_{0}}$ is choosen so
that $\Delta _{0}\protect\beta _{0}=15$.}
\end{figure}

It is clearly seen from Fig.3 that the CVC has an ``S-shaped'' form. This
type of the CVC may also occur in semiconductors \cite{VK} as a result of
the so-called overheating mechanism and in bulk superconductors in the
presence of a dissipative current \cite{Elesin,SchonRev,Dynes,Al'tshuler}.
This phenomenon is possible for a special dependence of the heat source $%
J(V,T)$ and heat absorption term $S(T)$ on the effective temperature $T$ and
applied voltage $V.$ The ``N-shaped'' CVC may also arise as a result of the
overheating mechanism \cite{Tinkham,VKJETP,Keizer,Mints,Vodolazov}.

One can show in the same way as it was done in Ref. \cite{VK} that the
states corresponding to the part of the CVC with a negative differential
resistance are unstable and, in the case of a fixed total current \cite{Foot}%
, the system is stratified into layers with different effective temperatures
and current densities. The form of the current filaments can be found from
Eq. (\ref{EqTeff}).

We introduce a new effective ``temperature'' $\vartheta =\int_{1}^{\tilde{T}%
}d\tilde{T}_{1}M(\tilde{T}_{1})\tilde{T}_{1}$ and a function of this
``temperature'' $W(\vartheta )=\int_{\mathbf{0}}^{\vartheta }[J(V,\tilde{T}%
(\vartheta _{1})-S(\tilde{T}(\vartheta _{1}))]d\vartheta _{1}.$ Integrating
Eq. (\ref{EqTeff}) over the temperature $\tilde{T}$ and then over $\vartheta
$\ we arrive at the equation

\begin{equation}
(1/2)l_{0}^{2}(\partial \vartheta /\partial x)^{2}=W_{0}-W(\vartheta ,V)
\label{Energy}
\end{equation}%
where $l_{0}^{-1}=\kappa _{r}=\kappa _{l}.$

Eq. (\ref{Energy}) is already quite simple and its solution $\vartheta $
describes the dependence of the temperature $\tilde{T}$ on the coordinate $x$
along the interface. One can rather easily see that this dependence can be
non-trivial. This originates from the fact that in a narrow interval of $V$,
where the CVC is a multi-valued function, the function $W(\vartheta ,V)$ has
three extrema at $\vartheta _{k}(T_{k})$, where $k=1,2,3$ (see Fig.4a) As a
consequence, the solutions of Eq. (\ref{Energy}), $l_{0}(\partial \vartheta
/\partial x)=\sqrt{2}\sqrt{W_{0}-W(\vartheta ,V)}$, have different forms:
solitons (instantons), oscillatory temperature distributions or domain walls.

In Fig.4b we show qualitatively phase trajectories illustrating this
conclusion. At a certain voltage $V_{0}$ the function $W(\vartheta )$ has
the same values at the maxima: $W(\vartheta _{1})=W(\vartheta _{3}).$ This
means that the voltage $V_{0}$ satisfies the condition: $\int_{\vartheta
_{1}}^{\vartheta _{3}}[J(V_{0},\tilde{T}(\theta )-S(\tilde{T}(\theta
))]d\theta =0.$ The trajectory connecting these maxima (separatrix)
corresponds to a solution like a domain wall: $l_{0}(\partial \vartheta
/\partial x)=\sqrt{2}\sqrt{W(\vartheta _{1},V_{0})-W(\vartheta ,V_{0})}.$
The spatial distribution of the effective temperature $T$ related to this
trajectory is a decay of $T$ from the value $T_{3}$ to $T_{1}$ over a length
of the order of $l_{0}$. The trajectories close to the separatrix describe
the domain structure (see Fig.4c). This structure is analogous to the one in
a superconductor with an exchange field in the LOFF state.

\begin{figure}[tbp]
\begin{center}
\includegraphics[width=8cm, height=8cm]{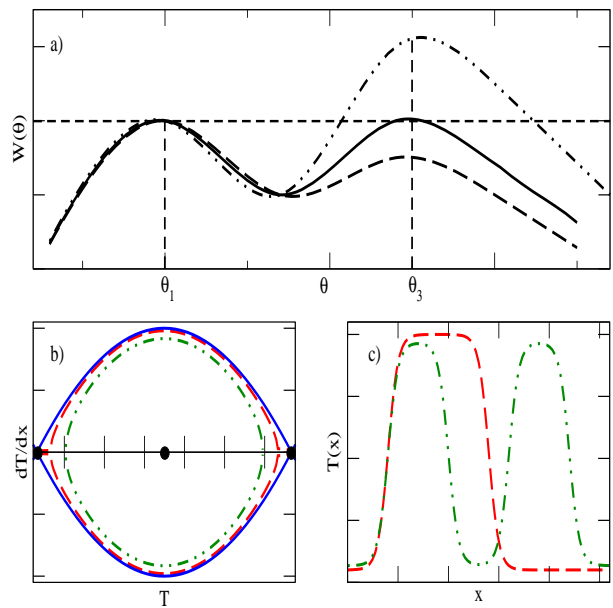}
\end{center}
\caption{a) The potential $W(\protect\theta)$ for different $V$: $V = V_{0}$
(solid line); $V < V_{0}$ (dashed line) and $V > V_{0}$ (dash-dotted line).
b) Phase trajectories ($T,dT/dx$). Solid line shows a separatrix (a domain
wall in $T(x)$ distribution); dashed and dotted lines show domain-like
structure of $T(x)$ distribution; c) Spatial dependence of the effective
temperature corresponding to a domain-like structure (arbitrary units)}
\end{figure}

At the same time, the inhomogeneous state in the SN systems out of the
equilibrium differs from the LOFF state. In the latter case the solution of
the self-consistency equation should correspond to the minimum of the free
energy whereas in our case the choice of the solution is dictated by the
bias current $I_{b}$ and by the stability against\ small perturbations. One
can show that the most stable solution has the minimal number of zeros of
the function $\partial T(x)/\partial x$ \cite{KPet,VK}.

This fact suggests the following scenario:

a) With increasing the bias current $I_{b}$ the dependence $I(V)$ follows
the CVC for the homogeneous case.

b) At $I_{b}>I_{m}$ the voltage $V$ jumps to a lower value close to $V_{0}$
and remains almost constant with increasing $I_{b}$ (see Fig.3a and \cite{VK}%
)$.$

c) When the current $I_{b}$ exceeds the value $I_{3}$, the voltage $V$
increases along the CVC for the homogeneous case. If $I_{b}$ decreases from $%
I_{3}$, the voltage decreases and a jump to a higher voltage close to $V_{0}$
occurs at $I_{b}\leq I_{M}$ (a hysteretic behavior).

It would be interesting to study experimentally the inhomogeneous
distribution of the effective temperature $T(x)$, current and the gap $%
\Delta (x).$ This could be done, for example, by applying the point-contact
spectroscopy to N/S systems (see a possible setup in Fig.1b). The
distribution function $f_{+}$ in these systems has the same form as in N/S/N
systems. Therefore, by measuring the spatial distribution $\Delta (x)$, one
can get information about the inhomogeneous states in the N/S systems.

It is interesting to note that the ideas that the symmetry breaking of the
LOFF type may occur not only in the superconductors but also in the quantum
chromo-dynamics, astrophysics and in cold gases (see the review \cite%
{Nardulli} and references therein) have been put forward recently. Many
theoretical papers have been published since the pioneering works \cite%
{FF,LO}, in which different aspects of the LOFF state in superconductors and
conditions necessary for this state were analyzed (see Refs. \cite%
{Nardulli,Mineev} and references therein). As to the experimental
observation of the inhomogeneous states predicted, the situation is not as
clear. Although some observations in low dimensional superconductors and
superconductors with heavy fermions can be interpreted in terms of the LOFF
states, there are no convincing evidences in favor of such states in
ordinary $s$-wave superconductors.

So, the investigation of non-equilibrium superconductors, for example
superconductors in tunnel N/S/N systems, can provide one more possibility to
observe the LOFF state experimentally. Such experiments may be useful for
understanding the LOFF states and the conditions under which they can be
realized. We mention here another interesting example of the analogy between
a non-equilibrium superconducting system and equilibrium superconducting
system with an exchange field, namely, the $\pi -$state may arise not only
in S/F/S junctions \cite{GolubovRMP,BuzdinRMP,BVERMP}, but also in S/N/S
junctions with a non-equilibrium distribution function \cite%
{KlapWees,VolkovPRL,SchonPRL,Yip}(here F denotes a ferromagnetic layer).

\bigskip

\section{Conclusions}

In conclusion, using the approximation of an effective temperature of
quasi-particles, we have studied non-equilibrium states in a tunnel NSN
structure with a low barrier transparency. It is found that for certain
values of parameters, the CVC of the system may have an S-shaped form. The
uniform state corresponding to the part of the CVC with negative
differential conductance is unstable, and therefore a nonuniform current $%
I(x)$ and temperature $T(x)$ distribution is established in the system with
a fixed total current. We discuss the analogy with the nonuniform LOFF
states in equilibrium superconductors and the possibilities of experimental
observation of the nonuniform states in SN structures.

We would like to thank SFB 491 for financial support.


\begin{thebibliography}{99}
\bibitem{LambRaim} C.J.\ Lambert and R.\ Raimondi, J. Phys.: Condens. Matter
\textbf{10}, 901 (1998).

\bibitem{Been} C.W.J.\ Beenakker, in Mesoscopic Quantum Physics, ed. by E.
Akkermans; G.\ Montambaux, J.-L. Pichard, and J. Zinn-Justin,
(North-Holland, Amsterdam, 1995).

\bibitem{ZaikinRev} W.\ Belzig, G. Sch\"{o}n, C. Bruder, and A.D. Zaikin,
Superlattices and Microstructures \textbf{25}, 1251 (1999).

\bibitem{NazRev} Yu.V.\ Nazarov, Superlattices and Microstructures \textbf{25%
}, 1221 (1999).

\bibitem{AVZ} S.N. Artemenko, A.F. Volkov and A.V. Zaitsev, Solid State
Commun. \textbf{30}, 771 (1979).

\bibitem{Marg} V.N. Gubankov and N.M. Margolin, JETP Lett. \textbf{29}, 673
(1979).

\bibitem{NazarovPRL} Yu.V.\ Nazarov and T.H.\ Stoof, Phys. Rev. Lett.
\textbf{76}, 823 (1996).

\bibitem{LambVolkov} A.F.\ Volkov, N. Allsopp, and C.J. Lambert, J. Phys.\
Condens Matter \textbf{8}, 45 (1996).

\bibitem{Pannetier} P. Charlat, H. Courtois, Ph. Gandit, D. Mailly, A. F.
Volkov, and B. Pannetier, Phys. Rev. Lett. \textbf{77}, 4950 (1996).

\bibitem{Elesin} V.M.Galitskii, V.F.Elesin, and Yu.V.Kopaev, in
Nonequilibrium Superconductivity, edited by D.N. Langenberg and A.I. Larkin
(Elsevier, Amsterdam, 1986), p. 377.

\bibitem{SchonRev} G. Sch\"{o}n, Physica B\&C, \textbf{109}, 1677 (1982).

\bibitem{Mints} A. Vl. Gurevich and R. G. Mints, Rev. Mod. Phys. \textbf{59}%
, 941 (1987).

\bibitem{Eliashberg} G.M. Eliashberg, JETP Lett. \textbf{11}, 114 (1970).

\bibitem{Ivlev} G.M. Eliashberg and B.I. Ivlev, in Nonequilibrium
Superconductivity, edited by D.N. Langenberg and A.I. Larkin (Elsevier,
Amsterdam, 1986), p. 211.

\bibitem{KlapWees} J. J. Baselmans, A. F. Morpurgo, T. M. Klapwijk, and B.
J. van Wees, Nature \textbf{397}, 43 (1999).

\bibitem{VolkovPRL} A.F.\ Volkov, Phys. Rev. Lett. \textbf{74}, 4730 (1995).

\bibitem{SchonPRL} F. K. Wilhelm, G. Sch\"{o}n, and A. D. Zaikin, Phys. Rev.
Lett. \textbf{81}, 1682 (1998).

\bibitem{Yip} S. K. Yip, Phys. Rev. B \textbf{58}, 5803 (1998).

\bibitem{Pekola} F. Giazotto, T. T. Heikkila, A. Luukanen, A. M. Savin, J.
P. Pekola, Rev. Mod. Phys. \textbf{78}, 217 (2006).

\bibitem{VZK} A.F.Volkov, A.V. Zaitsev, and T.M. Klapwijk, Physica \textbf{C
210}, 21 (1993); A.V.Zaitsev, A. F. Volkov, S. W. D. Bailey, and C. J.
Lambert, Phys.Rev. B \textbf{60}, 3559 (1999).

\bibitem{Keizer} R.S. Keizer, M.G.\ Flokstra, J. Aarts, and T.M. Klapwijk,
Phys. Rev. Lett. \textbf{96}, 147002 (2006).

\bibitem{Nazarov} I. Snyman, Yu. V. Nazarov, Phys. Rev. B \textbf{79},
014510 (2009).

\bibitem{Tinkham} W.J. Skocpol, M.R. Beasley, and M. Tinkham, J. Appl. Phys.
\textbf{45},4054 (1974).

\bibitem{VKJETP} A.F. Volkov and Sh.M. Kogan, JETP\ Letters \textbf{19}, 4
(1974).

\bibitem{Vodolazov} D.Y. Vodolazov, F.M. Peeters, L. Piraux, S.
Matefi-Tempfli, S. Michotte, Phys. Rev. Lett. \textbf{91}, 157001 (2003).

\bibitem{Dynes} R. C. Dynes, V. Narayanamurti, and J. P. Garno, Phys. Rev.
Lett. \textbf{39}, 229 (1977).

\bibitem{Gray} K.E. Gray and H. W. Willemsen, J. Low Temp. Phys. \textbf{31}%
, 911 (1978).

\bibitem{India} M. Ovadia, B. Sac\'{e}p\'{e}, and D. Shahar, Phys. Rev.
Lett. \textbf{102}, 176802 (2009).

\bibitem{Al'tshuler} B.L. Altshuler, V.E. Kravtsov, I.V. Lerner, I.L.
Aleiner, Phys. Rev. Lett. \textbf{102}, 176803 (2009).

\bibitem{Readley} B.K.\ Ridley, Proc. Phys. Soc. \textbf{82}, 954 (1963).

\bibitem{VK} A.F. Volkov and Sh.M. Kogan, Sov.Phys.Usp. \textbf{11}, 881
(1969).

\bibitem{Schmid} A. Schmid and G. Sch\"{o}n, J. Low Temp. Phys. \textbf{20},
207 (1975).

\bibitem{FF} P. Fulde, and R. A. Ferrell, Phys. Rev. \textbf{135}, A550
(1964).

\bibitem{LO} A. J. Larkin and Y. N. Ovchinnikov, Zh. Eksp. Teor. Fiz.
\textbf{47}, 1136 (1964) [Sov. Phys. JETP \textbf{20}, 762 (1965)].

\bibitem{Langenberg} S.B. Kaplan, C.C.\ Chi, D.N. Langenberg, J.J.\ Chang,
S. Jafarey and D.J. Scalapino, Phys. Rev. B \textbf{14}, 4854 (1976).

\bibitem{Kopnin} N. B. Kopnin, \textit{Theory of Nonequilibrium
Superconductivity}, Clarendon Press, Oxford (2001).

\bibitem{Al'tshuler1} B.L. Altshuler, Zh.Eksp.Teor. Fiz. \textbf{75}, 1330
(1978) [Sov. Phys. JETP \textbf{48}, 670 (1978)].

\bibitem{Reizer} M.Yu. Reizer and A.V.\ Sergeev, Zh.Eksp.Teor. Fiz. \textbf{%
90}, 1056 (1986) [Sov. Phys. JETP \textbf{63}, 616 (1986)].

\bibitem{Sergeev} A.V. Sergeev and V. Mitin, Phys. Rev. B \textbf{61}, 6041
(2000).

\bibitem{Parker} W.H. Parker, Phys. Rev. B \textbf{12}, 3667 (1975).

\bibitem{Hekking} A. S. Vasenko, F. W. J. Hekking, J. Low Temp. Phys.
\textbf{154}, 221 (2009).

\bibitem{SV} R. Seviour and A. F. Volkov, Phys. Rev. B \textbf{62}, R6116
(2000); V. R. Kogan, V. V. Pavlovskii, and A. F. Volkov, Europhys. Lett.
\textbf{59}, 875 (2002).

\bibitem{VK73} A. F. Volkov and S. M. Kogan, Sov. Phys. JETP \textbf{38}%
,1018 (1974).

\bibitem{Foot1} According to a recent paper \cite{Kopnin1} this
approximation is violated if the effective temperature $T$ is much less than
$eV$. In our case $T\gtrsim eV$.

\bibitem{Kopnin1} N.B. Kopnin et al., cond-mat/0905.1210 (unpublished).

\bibitem{Foot} If the applied voltage is fixed (a voltage-biased case), the
system jumps from the low-conductance to the high-conductance state and vice
versa (see, for example, \cite{VK,Al'tshuler}).

\bibitem{KPet} B.W. Knight and G.A. Peterson, Phys. Rev. \textbf{155}, 393
(1967).

\bibitem{Nardulli} R. Casalbuoni and G. Nardulli, Rev. Mod. Phys. \textbf{76}%
, 263 (2004);

\bibitem{Mineev} M. Houzet, V. P. Mineev, Phys. Rev. B \textbf{76}, 224508
(2007), Y. Yanase, M. Sigrist, Journal of Physics: Conference Series \textbf{%
150}, 052287 (2009); D. Denisov, A. Buzdin, H. Shimahara, Phys. Rev. B
\textbf{79}, 064506 (2009).

\bibitem{GolubovRMP} A.A. Golubov, M.Y. Kupriyanov, and E.Il'ichev, Rev.
Mod. Phys. \textbf{76, }411 (2004);

\bibitem{BuzdinRMP} A. Buzdin, Rev. Mod. Phys. \textbf{77}, 935 (2005);

\bibitem{BVERMP} F. S. Bergeret, A. F. Volkov, K. B. Efetov, Rev. Mod. Phys.
\textbf{77, }1321 (2005).

\end{thebibliography}
\end{document}